\documentclass[12pt,preprint]{aastex}

\begin{document}
\slugcomment{Accepted by ApJS 2008 Jan 9; submitted 2007 Dec 3}
\title{Probing Interstellar Dust With Space-Based Coronagraphs}

\author{N.~J.~Turner, K.~Grogan and J.~B.~Breckinridge} \affil{Jet
  Propulsion Laboratory, California Institute of Technology, Pasadena,
  California 91109, USA; Neal.Turner@jpl.nasa.gov}

\begin{abstract}
  We show that space-based telescopes such as the proposed Terrestrial
  Planet Finder Coronagraph will be able to detect the light scattered
  by the interstellar grains along lines of sight passing near stars
  in our Galaxy.  The relative flux of the scattered light within one
  arcsecond of a star at 100~pc in a uniform interstellar medium of
  0.1 H~atoms cm$^{-3}$ is about $10^{-7}$.  The halo increases in
  strength with the distance to the star and is unlikely to limit the
  coronagraphic detection of planets around the nearest stars.  Grains
  passing within 100~AU of Sun-like stars are deflected by radiation,
  gravity and magnetic forces, leading to features in the scattered
  light that can potentially reveal the strength of the stellar wind,
  the orientation of the stellar magnetic field and the relative
  motion between the star and the surrounding interstellar medium.
\end{abstract}

\keywords{ISM: dust --- scattering --- radiative transfer ---
  techniques: high angular resolution}

\section{INTRODUCTION\label{sec:intro}}

Scattered starlight is a valuable tool for measuring the properties of
interstellar dust.  In optical reflection nebulae
\citep{wp92,c95,b02}, dark clouds \citep{fs76,wo90} and the diffuse
Galactic light \citep{hb91,mh95,wf97} the scattering grains typically
lie many arcseconds from the illuminating sources.  The dust can be
nearer to or further from us than the source and the detected photons
have been deflected through a wide range of angles.  Further
information about interstellar grains comes from stellar X-rays
scattered by foreground dust within a few arcseconds of the star
\citep{m95,sd98,ws01,dt03}.  The measurements sample the whole column
of material between us and the star because the grains are strongly
forward scattering at X-ray energies.

Here we show that high-contrast imaging with a space-based telescope
such as the proposed Terrestrial Planet Finder Coronagraph enables a
new kind of measurement for probing dust at projected separations
similar to the X-ray measurements.  The technique samples a range of
scattering angles because at visible wavelengths the scattering is
only mildly biased in the forward direction.  As a consequence, most
of the light comes from foreground grains lying close to the star.
The scattered light halo carries information on the interstellar dust
passing through the cavity opened in the interstellar gas by the
stellar wind.

Evidence for the existence of interstellar grains within our own
heliosphere was obtained by the Ulysses spacecraft when 55 dust
impacts as measured by the Cosmic Dust Experiment were identified as
interstellar by their speed, mass and arrival direction \citep{gg94}.
The overall contribution of interstellar dust to the zodiacal cloud
--- the debris disk of the solar system --- is unknown, although the
fraction is presumably greater in the outer solar system.  Collisional
debris from the asteroid belt dominates inside 3~AU \citep{gd01}.
However the interstellar grains, charged by photoionization and
strongly influenced by the solar gravity, radiation pressure and
magnetic field, are able to penetrate deep into the inner solar
system.  A uniform incoming spatial distribution of interstellar
grains becomes strongly clumped as a function of particle size and
phase of the solar cycle, although the contribution of this component
to the all-sky thermal emission as viewed from 1~AU is negligible due
to the local dominance of asteroidal and cometary material
\citep{gd96}.

Starlight scattered by interstellar dust also is a source of noise in
coronagraphic planet searches.  We calculate the expected brightness
and distribution of the scattered light and show that the halo is
unlikely to affect the direct detection of Earth-like planets around
the nearest stars.  We compute the trajectories of interstellar grains
under the stellar gravity, radiation and magnetic forces and show that
the stellar wind produces observable signatures in the scattered
light.  For a large number of stars at distances greater than 100~pc,
high-contrast imaging can potentially yield detailed information about
the stellar wind and the adjacent interstellar medium.

The remaining five sections of the paper cover the radiative transfer
method (\S2), the results for uniformly-distributed dust, obtained
analytically assuming isotropic scattering (\S3) and numerically
including the anisotropy (\S4), the results for dust passing through a
model stellar wind (\S5), and a summary and conclusions (\S6).

\section{RADIATIVE TRANSFER\label{sec:rt}}

\begin{figure}[tb!]
  \epsscale{1.0}
  \plotone{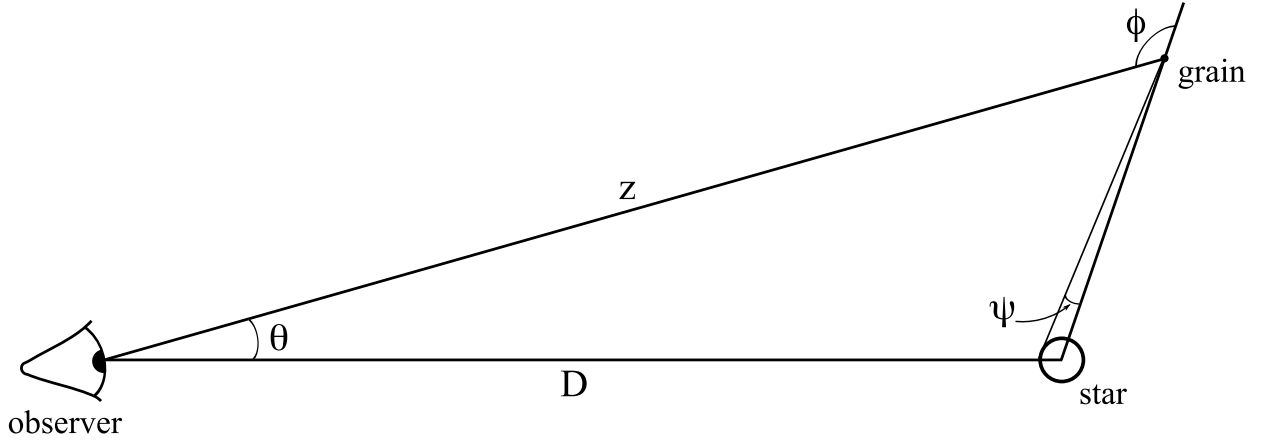}

  \figcaption{\sf Sketch of the scattering geometry for a star at
    distance $D$.  A representative interstellar grain lies along a
    line of sight offset from the star by an angle $\theta$ and is at
    a distance $z$.  The star subtends an angle $2\psi$ as seen from
    the grain.  Stellar photons reach our line of sight on scattering
    through an angle $\phi$.
    \label{fig:sketch}}
\end{figure}

The transfer of visible light through the optically-thin interstellar
medium is dominated by scattering and the radiative transfer equation
at wavelength $\lambda$ is
\begin{equation}\label{rt}
{dI_\lambda\over d\tau_\lambda} =
- {\omega_\lambda}
  \oint \Phi_\lambda({\bf\Omega}^\prime\rightarrow{\bf\Omega})
  I_\lambda({\bf\Omega}^\prime) d{\bf\Omega}^\prime,
\end{equation}
where $I_\lambda$ is the monochromatic specific intensity,
$\tau_\lambda$ the optical depth, $\omega_\lambda$ the albedo of the
scattering grains and $\Phi_\lambda$ the phase function describing
scattering from all directions ${\bf\Omega}^\prime$ into the beam
direction ${\bf\Omega}$.  The star is approximated by a disk of
uniform specific intensity $I^*_\lambda$, subtending a small angle
$2\psi$ as seen from a dust grain on our line of sight
(figure~\ref{fig:sketch}).  The scattering phase function is taken to
be constant across the face of the star, and the solid angle filled by
the star is $\pi\psi^2$.  The transfer equation reduces to
\begin{equation}\label{eqn:halort}
{dI_\lambda\over d\tau_\lambda} =
- {\omega_\lambda} \Phi_\lambda(\phi) \pi\psi^2 I^*_\lambda,
\end{equation}
where $\phi$ is the scattering angle between the rays joining star to
grain and grain to observer.  The intensity of the scattered light is
calculated by integrating eq.~\ref{eqn:halort} along the line of
sight.  The angles $\phi$ and $\psi$ vary with the distance from the
observer, $z$.  Writing the distance in place of the optical depth
using $d\tau_\lambda = \chi_\lambda\rho\,dz$, neglecting emission from
any distant background, and taking the simple case of an interstellar
medium with uniform opacity, the intensity is
\begin{equation}\label{eqn:si}
I_\lambda = \chi_\lambda\rho \omega_\lambda I_\lambda^* 
\int_0^\infty \Phi_\lambda(\phi) \pi\psi^2 dz.
\end{equation}

\section{ISOTROPIC SCATTERING\label{sec:isotropic}}

We first solve equation~\ref{eqn:si} assuming isotropic scattering,
with $\Phi_\lambda = 1/(4\pi)$ independent of $\phi$ and $z$.  Along a
line of sight offset by a small angle $\theta$ from the direction to
the star, the stellar angular radius $\psi$ varies as
\begin{equation}
\psi(z)={r_*/D\over\left(\theta^2+\left[1-z/D\right]^2\right)^{1/2}},
\end{equation}
where the star has radius $r_*$ and lies at distance $D$.  With a
change of variable to $y=(1-z/D)/\theta$, the solution to the transfer
equation is
\begin{equation}
{I_\lambda(\theta)\over I_\lambda^*} =
  {\chi_\lambda\rho \omega_\lambda r_*^2\over 4D\theta}
  \int_{-\infty}^{1/\theta} {dy\over 1+y^2}.
\end{equation}
The integral is very nearly $\pi$ because $1/\theta$ is almost
$+\infty$ and the integrand falls off rapidly with $y$.

As a specific case we take the Sun observed from a distance of 100~pc
at wavelength 547~nm through a uniform interstellar medium with 0.1
H~atoms cm$^{-3}$, similar to the densities found in the solar
neighborhood \citep{bk82,sf02}.  For Milky Way grains, the total
opacity $\chi_\lambda\rho = 4.9\times 10^{-22}$ cm$^2$ H$^{-1} \times
0.1$ H~cm$^{-3} = 4.9\times 10^{-23}$ cm$^{-1}$ and albedo
$\omega_\lambda=0.66$ \citep{d03}.  The optical depth to the star
$\chi_\lambda\rho D\approx 10^{-2}$.  The predicted surface brightness
ratio at an angular separation $\theta$ is
\begin{equation}\label{eqn:isotropic}
{I_\lambda(\theta)\over I_\lambda^*} =
  {\pi\chi_\lambda\rho \omega_\lambda r_*^2\over 4D\theta} =
  8\times 10^{-16}
  \left({n_H\over 0.1\,{\rm cm}^{-3}}\right)
  \left({100\,{\rm pc}\over D}\right)
  \left({0.1^{\prime\prime}\over\theta}\right)
\end{equation}
and falls off inversely with the separation.  The integrated flux
within an angle $\theta_{max}$ is proportional to the optical depth,
\begin{equation}
{F_\lambda(\theta_{max})\over F_\lambda^*} = 8\times 10^{-8}
\left({n_H\over 0.1\,{\rm cm}^{-3}}\right)
\left({D\over 100\,{\rm pc}}\right)
\left({\theta_{max}\over 1^{\prime\prime}}\right)
\end{equation}
and the flux within one arcsecond is eighty times greater than the
nominal threshold $10^{-9} F_\lambda^*$ for detection by the proposed
Terrestrial Planet Finder Coronagraph.

\section{ANISOTROPIC SCATTERING\label{sec:anisotropic}}

\begin{figure}[tb!]
  \epsscale{0.5}
  \plotone{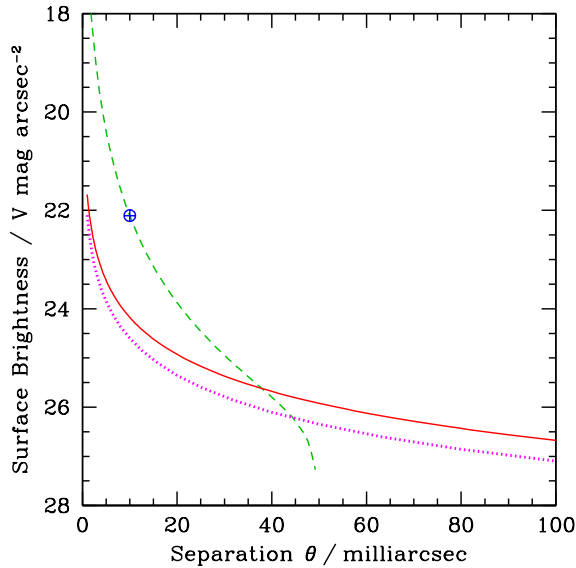}

  \figcaption{\sf Surface brightness versus separation from the star
    for a solar analog exozodiacal cloud (dashed curve) and
    uniformly-distributed interstellar grains, scattering either
    isotropically (dotted line) or anisotropically according to the
    \cite{d03} phase function (solid line).  A crossed circle marks
    the approximate angular separation and surface brightness of an
    exo-Earth at the nominal angular resolution of the Terrestrial
    Planet Finder Coronagraph \citep{ls06}.  The star lies at a
    distance of 100~pc.
    \label{fig:anisotropic}}
\end{figure}

Interstellar grains scatter visible light anisotropically.  We treat
this effect using a standard model for grains in the diffuse
interstellar medium of the Milky Way, with the phase function
$\Phi_\lambda(\phi)$ computed from Mie theory by \cite{d03}.  Forward
scattering is moderately preferred and at the chosen wavelength of
547~nm, each event has a 50\% chance of deflecting the photon through
an angle less than 43\arcdeg.  The interstellar medium is uniform with
the same density as in section~\ref{sec:isotropic}.  The transfer
equation~\ref{eqn:halort} is numerically integrated along the line of
sight to the observer using a fourth-order Runge-Kutta method.  We
have checked that the method reproduces the analytic solution given by
equation~\ref{eqn:isotropic} in the case of isotropic scattering.
Including the preferential forward scattering yields a halo 48\%
brighter.  The results for a solar analog star at a distance of 100~pc
are shown in figure~\ref{fig:anisotropic}.  Also shown for comparison
is a solar analog exozodiacal cloud from appendix 1.B of \cite{ls06}.
The light scattered from the interstellar grains is brighter than that
from the exozodiacal cloud outside 50~milliarcsec.  The exozodiacal
cloud is inclined 60\arcdeg\ from face-on and its brightness was
calculated including anisotropic scattering.

\begin{figure}[tb!]
  \epsscale{0.5}
  \plotone{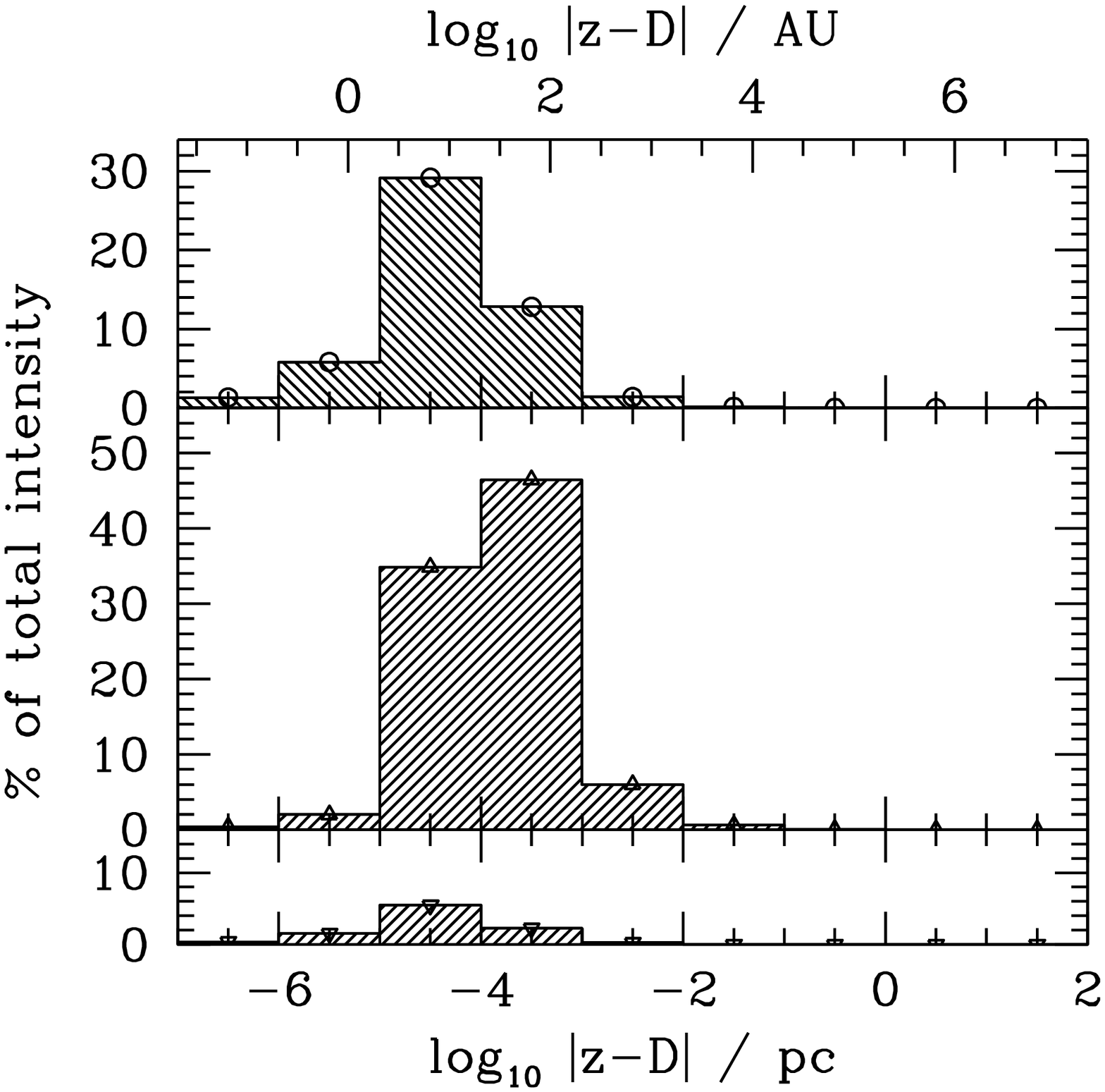}

  \figcaption{\sf Contribution to the surface brightness from grains
    at different locations along the line of sight passing 100~mas or
    10~AU from the star in figure~\ref{fig:anisotropic}.  The three
    panels show isotropic scattering (top) and anisotropic scattering
    by grains in front of (center) and behind the plane containing the
    star (bottom).  The horizontal axes give the distance from the
    star plane in AU (above the chart) and parsecs (below).  Most of
    the light is scattered into our line of sight by grains within
    100~AU of the star.
    \label{fig:where}}
\end{figure}

Much of the interstellar scattered light comes from grains passing
close by the star, due to the inverse square falloff of the stellar
illumination.  Most comes from grains whose distance from the star is
less than ten times the projected separation in the plane of the sky.
The preferential forward scattering has a smaller effect, making the
dust in front of the star contribute more than the dust at the same
distance behind.  The contributions from dust at different positions
along a line of sight passing 0.1~arcsecond from a star at 100~pc are
shown in figure~\ref{fig:where}.  Overall, the scattered light halo is
most sensitive to the distribution of interstellar dust in the
vicinity of the illuminating star.

Measurements and models of the optical properties of the dust in the
diffuse interstellar medium allow a range of values for the albedo and
the degree of forward scattering \citep{g04}.  The \cite{d03} model
used above lies near the middle of the range in both the albedo and
the scattering asymmetry as measured by the mean cosine scattering
angle $g=\left<\cos\phi\right>$ at 547~nm.  The appearance of the halo
is qualitatively independent of the dust properties over the allowed
range.  The halo brightness is simply proportional to the albedo
(equation~\ref{eqn:isotropic}).  The scattering asymmetry affects both
the halo brightness and the location of the grains giving most of the
light.  We examine the dependence by re-computing the halo using a
standard Henyey-Greenstein phase function with asymmetry parameters up
to the maximum $g=0.8$ consistent with the range of the measurements.
Half the photons are deflected through angles less than 19\arcdeg.
The resulting halo is 2.19~times brighter than with the \cite{d03}
phase function.  The fraction of the light 0.1~arcsecond from the star
that is scattered by grains lying more than 200~AU from the star rises
from 7\% to 21\%.  A star surrounded by the more forward-throwing
grains has a brighter halo with a slightly lower contrast for any
features arising from dust passing close by the star.

\section{PROBING ASTROSPHERES\label{sec:astrospheres}}

Sub-micron interstellar grains entering the solar system are deflected
by the solar gravity, radiation pressure and solar wind
electromagnetic forces \citep{gm79}.  The highest densities of grains
0.1~micron in radius occur where the particles pile up against the
solar wind and where the magnetic forces concentrate particles over
the solar magnetic poles \citep{gd96,l00}.  In this section we explore
whether similar effects can be detected in the scattered light halos
around other stars.

\subsection{Stellar Wind Model}

The star is taken to be a solar analog and its wind is modeled
following \cite{gm79}, \cite{gd96} and \cite{l00} by applying the
expanding corona model of the solar wind magnetic field \citep{p58}.
The gas streaming outward from the star draws out the stellar magnetic
field lines so that near the star the field is approximately radial.
The more distant field lines are dragged into the shape of an
Archimedean spiral as the star rotates.  The radial and azimuthal
components of the field are
\begin{equation}
B_r = B_0/r^2
\end{equation}
and
\begin{equation}
B_\phi = - B_0 \omega b^2(r-b) \sin a / V_{sw} r^2,
\end{equation}
where $\omega$ is the stellar angular velocity, $a$ is the
astrocentric colatitude angle, $V_{sw}$ is the stellar wind speed
(chosen as 400~km~s$^{-1}$) and $b$ is the distance at which the
reference field strength $B_0$ is taken.  The Lorentz force is then
\begin{equation}\label{eqn:lorentz}
\vec{F}_L = \frac{q}{c} [(\vec{V}_g \times \vec{B}) - (\vec{V}_{sw} 
\times
\vec{B})]
\end{equation}
where $q$ is the grain charge and $V_g$ the grain speed. The grain
charge $q$ is given by
\begin{equation}
q = 4\pi \varepsilon_0 U s
\end{equation}
where $s$ is the grain radius, $\varepsilon_0 = 8.86\times
10^{-12}$~CV$^{-1}$m$^{-1}$ is the permittivity of free space and $U$
is the grain surface potential.  Dust in the stellar wind cavity is
expected to be positively charged due to the dominance of stellar
photoionization over the competing mechanism of the `sticking' of
stellar wind electrons \citep{m81}.  Since the stellar wind density
roughly follows an inverse square law decrease with astrocentric
distance, particles are charged to +5-10~volts with little distance
variation.  We adopt a surface potential $U$ of +5~volts.  Notice that
the dominant electric part in equation~\ref{eqn:lorentz} is
proportional to $B_\phi$ and therefore to $r^{-1}$.  This can make the
instantaneous Lorentz force dominate on small grains at large
astrocentric distances.  The ballerina model proposed by \cite{a77}
gives an explanation for solar system spacecraft measurements which
indicate that near the ecliptic plane the interplanetary magnetic
field is directed inward in certain regions and outward in others.  In
this picture a current sheet separates plasma from either hemisphere
carrying fields of opposite polarity.  As the sheet rotates with the
Sun the small up and down displacements in the sheet, similar to the
wave motion of the skirt of a spinning ballerina, explain the observed
effect.  We use a simplified model of this phenomenon to describe the
magnetic field strength averaged over a stellar rotation by the
inclusion of a current sheet which changes its tilt angle
$\varepsilon$ to the stellar equatorial plane at a constant angular
velocity of $(1/11)\pi$ radians per year over the stellar magnetic
activity cycle \citep{g94}, so that the sheet completes one
360\arcdeg\ rotation in 22~years.  The averaged field at latitude $A$
is then a fraction $2\arcsin(\tan A/\tan\varepsilon)/\pi$ of the
unipolar field.  The line of nodes defined by the intersection of the
current sheet and stellar equatorial plane will rotate through
360\arcdeg\ in one 27~day stellar rotation.  The effect of radiation
pressure is to reduce the stellar gravitational force, the magnitude
of the radiation pressure being given by
\begin{equation}
P_{pr} = \frac{S}{c} Q_{pr}
\end{equation}
where $S$ is the flux, $c$ is the speed of light and $Q_{pr}$ is the
radiation pressure efficiency of the particle.

\subsection{Particle Properties}

Grains larger than one micron are rare in the diffuse interstellar
medium according to extinction curve fitting results
\citep{mr77,dl84,wd01}.  Grains smaller than 0.01~micron have little
optical cross-section and furthermore are largely excluded from the
heliosphere by magnetic forces due to their high charge-to-mass
ratios.  Grains in the 0.1~micron size range experience a radiation
pressure force larger than the gravitational force but nevertheless
are able to penetrate deep into the solar system \citep{gd96}.  Most
of the scattered light from interstellar grains passing through the
astrosphere of a solar analog star is likely due to grains about
0.1~micron in size.  We therefore concentrate on grains with radii
between 0.02 and 0.3~microns.

The ratio $\beta$ of the radiation pressure force to the gravitational
force may be expressed in terms of $Q_{pr}$ as
\begin{equation}
\beta = \frac{5.7\times 10^{-5} Q_{pr}}{s\rho},
\end{equation}
where $\rho$ is the density of the particle \citep{bl79}.  The optical
efficiencies of the grains are found from Mie theory.  We assume
astronomical silicate composition and optical constants \citep{ld93}.
Clearly, the particle spatial distributions will vary with the
particle size; for example, the gravitational force will dominate for
larger particles, whereas the interaction with the magnetic field will
become increasingly important at the small particle end of the size
range.  The distribution of particles will also depend on the stellar
magnetic cycle, so that the particle positions will represent a
`snapshot' in time rather than an equilibrium configuration.

\subsection{Dynamical Calculations}

Taking into account gravity, radiation pressure and the stellar
magnetic field model outlined above, we follow the trajectories of a
uniform distribution of dust particles at infinity as they approach
the star from the upstream direction at a speed of 26~km~s$^{-1}$.
The equations of motion are integrated at small intervals of time and
the position, speed and acceleration of the particles are calculated.
The local properties of the stellar wind and magnetic field are
updated as the integration progresses.  We continue the integration
until the heliosphere analog is filled with dust particles and stop
the run at a given phase of the stellar magnetic field cycle with the
magnetic field orientation matching that of the Sun in the year 1990.

\begin{figure}[tb!]
  \begin{center}
  \plotone{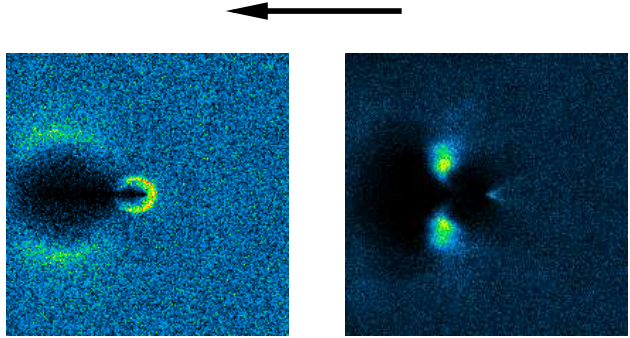}
  \end{center}
  \figcaption{\sf Interstellar dust density near a Sun-like star.  The
    star is centered in each image and the dust enters from the right,
    as shown by the arrow.  At left is a slice through the equatorial
    plane and at right is the polar plane containing the velocity
    vector.  The images are 400~AU on a side.  The color scale is
    linear between zero (black) and the maximum density (red).
    Focusing by the stellar gravity, radiation and magnetic forces
    enhances the density of the 0.1~$\mu$m particles up to 6~times at
    left and 15~times on the right.
    \label{fig:density}}
\end{figure}

The dust density is then calculated by binning the particles in cubes
2~AU on a side and normalizing to the upstream interstellar density.
The resulting distributions are broadly similar for particles 0.02 to
0.3~$\mu$m in radius.  The distribution of 0.1~$\mu$m grains is shown
in figure~\ref{fig:density}.  The particles are pushed by the magnetic
forces into the equatorial current sheet, and pile up 10~AU upstream
from the star (left panel), where the density reaches six times that
in the interstellar medium.  The greatest density enhancement of all
is a factor fifteen and occurs in two fins over the magnetic poles and
60~AU downstream from the star (right panel).  These result from the
focusing of the grains by the inclined magnetic fields off the
equatorial plane.

\subsection{Synthetic Images}

\begin{figure}[tb!]
  \begin{center}
  \plotone{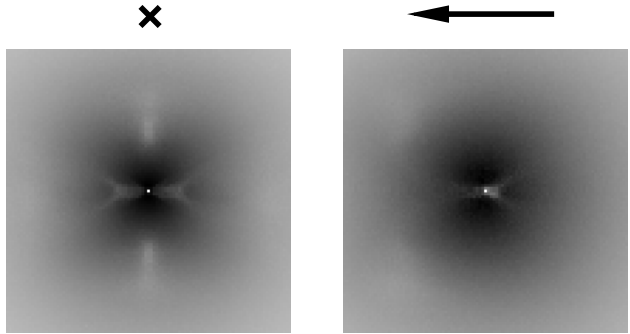}
  \end{center}
  \figcaption{\sf Model images of the starlight scattered by
    0.1~$\mu$m interstellar grains around a Sun-like star at a
    distance of 100~parsec.  The star is viewed looking upstream (left
    panel) and from the side (right panel).  The cross and arrow
    indicate the flow of the approaching grains.  The images are
    2~arcsec or 200~AU across, and were divided by the profile for
    uniformly-distributed dust.  Dark regions show a deficit of
    scattered light where magnetic and radiation forces exclude
    particles near the star.  The gray scale is linear between the
    minimum brightness ratio of about 0.1 (black) and the maximum of
    unity (white).
    \label{fig:halo}}
\end{figure}

\begin{figure}[tb!]
  \epsscale{0.5}
  \plotone{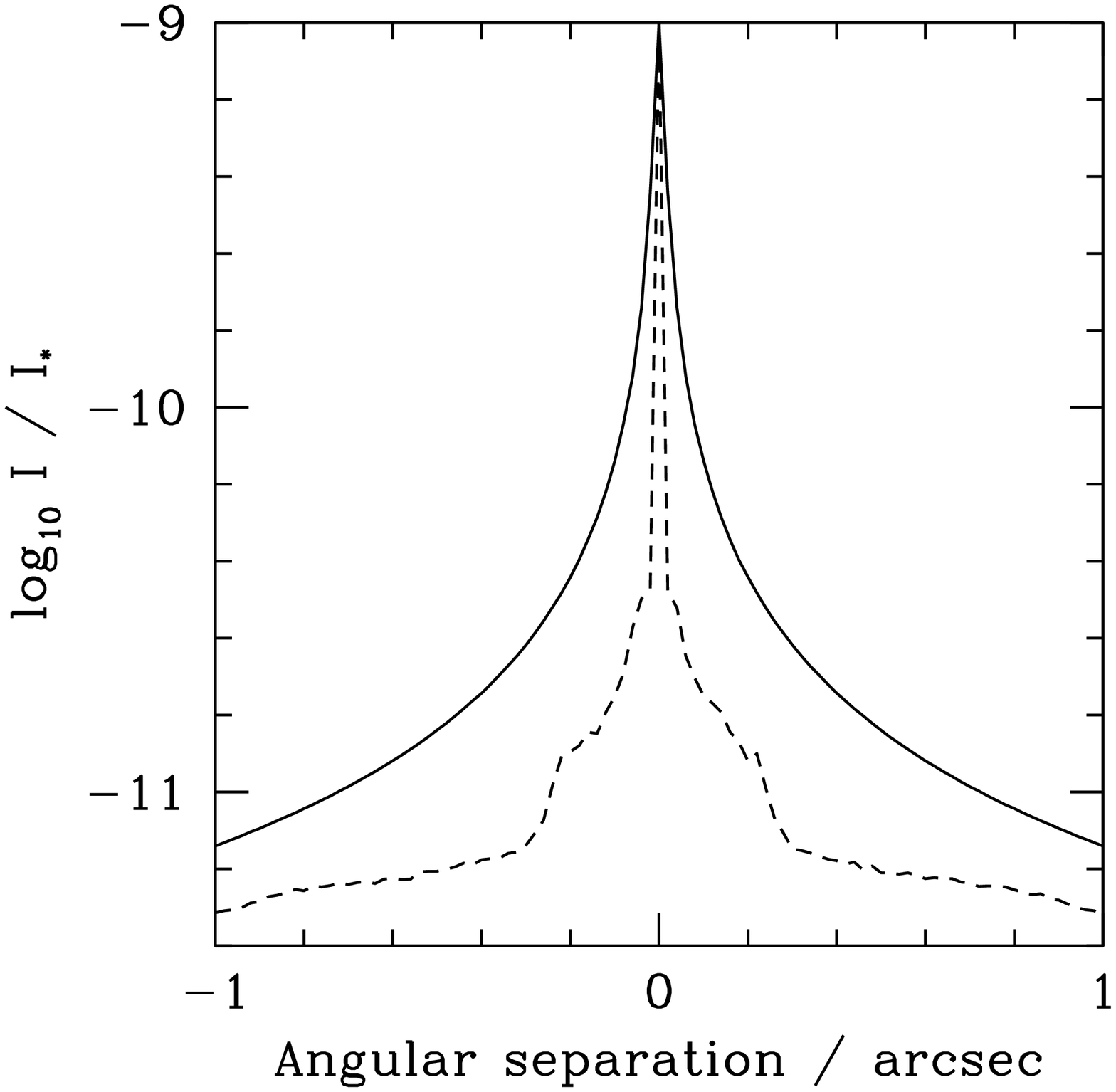}

  \figcaption{\sf Model profiles of the starlight scattered by
    0.1~$\mu$m interstellar grains around a Sun-like star at a
    distance of 100~parsec.  The solid line is for
    uniformly-distributed dust.  The dashed line shows the effects of
    stellar gravity, radiation and magnetic forces pushing aside dust
    near the star, and is a horizontal cut through the center of
    figure~\ref{fig:halo}, left panel.  The vertical axis is the count
    rate per pixel on a logarithmic scale, in units of the stellar
    count rate.  The pixels are 20~milliarcsec on a side.
    \label{fig:cut}}
\end{figure}

Synthetic images of the scattered starlight are constructed by solving
the transfer equation~\ref{eqn:halort} to find the intensity on a grid
of sky points, treating the anisotropic scattering as described in
section~\ref{sec:anisotropic} with the \cite{d03} Mie theory phase
function.  No exozodiacal cloud is included.  Selected results are
shown in figures~\ref{fig:halo} and~\ref{fig:cut}.  In addition to an
overall deficit of scattered light relative to the case of
uniformly-distributed dust, several features produced by the magnetic
field are clearly visible, including the density enhancements in the
equatorial disk and the polar fins.

\section{CONCLUSIONS\label{sec:conclusions}}

We have shown that stars appear surrounded by halos of light scattered
by interstellar dust.  The halos are bright enough for detection with
the proposed Terrestrial Planet Finder Coronagraph around stars at
distances of 100~pc and greater.  The halo reveals the distribution of
dust in the vicinity of the star, providing opportunities to probe the
interstellar medium and measure the properties of the stellar wind.
Particles of different sizes have different distributions, with larger
particles approaching the star most closely \citep{gd96}.  This
size-sorting can potentially be used to measure the distribution of
grain sizes at selected locations in the interstellar medium and to
resolve whether particles larger than one micron are present in
significant numbers, as suggested by dust impact detector data from
the Ulysses and Galileo spacecraft \citep{fd99,wd01}.  Since the
grains are deflected by stellar wind magnetic forces, the features in
the scattered light halo can be used to estimate the wind speed and
the relative motion between the star and the surrounding interstellar
medium.  The results will complement the wind mass loss rates
estimated using the Lyman-$\alpha$ absorption by the gas collected
near stellar wind bow shocks \citep{wm05}.  The scattering halos will
make it possible in addition to determine the wind magnetic field
strength and orientation (figure~\ref{fig:halo}) and, by observations
spanning a decade or so, the variation over the stellar activity cycle
\citep{l00}.  Imaging the halos will yield fresh information about the
winds of stars with a range of ages and could lead to a better
understanding of stellar spin histories \citep{bf97} and the long-term
evolution of stellar activity.

\acknowledgments

This work was carried out at the Jet Propulsion Laboratory, California
Institute of Technology with support from the internal Research and
Technology Development Fund Innovative Spontaneous Concepts Program.

\end{document}